\documentclass[twocolumn,showpacs,preprintnumbers,amsmath,amssymb]{revtex4}

\usepackage{graphicx}
\usepackage{dcolumn}
\usepackage{bm}

\begin{document}

\title{Bosons Confined in Optical Lattices  :\\the Numerical Renormalization Group revisited}

\author{Lode Pollet}

\email{Lode.Pollet@UGent.be}
\author{Stefan Rombouts}

\affiliation{
Subatomic and Radiation Physics Department\\
Proeftuinstraat 86, 9000 Gent, Belgium 
}
\author{Kris Heyde}
\affiliation{
Subatomic and Radiation Physics Department\\
Proeftuinstraat 86, 9000 Gent, Belgium 
}
\author{Jorge Dukelsky}

\affiliation{ 
CSIC, Serrano 123 \\
28006 Madrid, Spain
\\
}

\date{\today}

\begin{abstract}
A Bose-Hubbard model, describing bosons in a harmonic trap
with a superimposed optical lattice, is studied using
a fast and accurate variational technique (MF+NRG):
the Gutzwiller mean-field (MF) ansatz is combined
with a Numerical Renormalization Group (NRG) procedure 
in order to improve on both.
Results are presented for one, two and three dimensions,
with particular attention to the experimentally accessible
momentum distribution and possible satellite peaks in this
distribution.
In one dimension, a comparison is made with exact
results obtained using Stochastich Series Expansion.
\end{abstract}


\maketitle

\section{\label{sec:Intro}Introduction}

The recent experiments by Greiner \textit{et al.}~\cite{Greiner02} on bosons
confined in an optical lattice demonstrated the transition between a Mott
phase and a Superfluid phase (SF), as was first predicted by Jaksch 
\textit{et al.}~\cite{Jaksch98}. 
The experiments are adequately described by a single band 
Bose-Hubbard Hamiltonian with on-site repulsion only. This type of repulsion 
leads to two different phases. A Mott insulating phase can exist at
commensurate fillings, with a quantum phase transition to a superfluid as the
density is shifted or the interaction strength weakened. In the experiments
however, the quadratic confining potential adds a new term to the Hamiltonian
that cuts off any long range correlations, but Mott and superfluid regions can
still occur. The experiments led to a complete revival of interest in the
bosonic model, thanks to the unprecedented control over the physical
parameters compared to former realizations. 

Other experimental realizations of bosonic lattice systems include $^4$He on
graphite~\cite{Zimanyi94}, superconducting islands or grains connected by
Josephson junctions~\cite{vanOudenaarden96}. In this case, Cooper paired
fermions are considered as bosons, at least approximately. Recently, attempts
have been made to investigate at what scales the fermionic nature of paired
fermions can play a role~\cite{Rombouts02}, and the result is that for the
energy scales considered here, individual atoms can safely be described by a
one-boson operator $b^{\dagger}$, as expected.  

Since the work by Fisher \textit{et al.}~\cite{Fisher89}, the determination of
the ground state phase diagram of the Bose-Hubbard model with on-site
repulsion only has attracted a lot of attention. Analytic studies using
mean-field theory~\cite{Fisher89, Sheshadri94, vanoosten01} and renormalization group 
techniques~\cite{Fisher89} led to a deeper physical understanding of the
model. Strong coupling expansions~\cite{Freericks94, Elstner99} gave a
better quantitative picture, while quantum Monte-Carlo
simulations~\cite{Scalettar91, Bat92, Krauth91, Cha91, Kash96} were
carried out in one and two dimensions. The one-dimensional
case was recently investigated using the Density Matrix Renormalization Group
(DMRG)~\cite{Kuhner98, Kuhner99}, yielding the most accurate results at
present. Longer range interactions can cause charge density wave, stripe or
even supersolid order~\cite{Batrouni00, vanOtterlo94, Kuhner99}. Disorder and
impurities can have dramatic effects~\cite{Fisher89, Kane92, Giamarchi87,
  Scalettar91, Kuhner99, Prokofev03} and lead to even other phases. The model
with a quadratic confining potential~\cite{Jaksch98} has been addressed in one
dimension~\cite{Batrouni02} and for a small lattice in three
dimensions~\cite{Kashurnikov02}, using quantum Monte-Carlo methods.  

In view of the enormous success of DMRG~\cite{White92} in bosonic~\cite{Kuhner99}
and fermionic~\cite{White98} real-space lattice models in one dimension, DMRG
has been extended beyond these models, towards applications in metallic
grains~\cite{Dukelsky99}, quantum chemistry~\cite{White98b,  Legeza03} and
first attempts have even been undertaken  towards applications in nuclear 
physics~\cite{Dukelsky01}. Unfortunately, an exact DMRG study in all 3
dimensions of the Bose-Hubbard model is not feasible with current computer
power. DMRG was a substantial improvement~\cite{White92} on the older
Numerical Renormalization Group (NRG), which had only a poor reputation in
dealing with long-range interactions between fermions~\cite{Wilson75,
  Bray79}, but we found it useful for bosonic systems (see
also~\cite{Bulla03}).     

The basic philosophy of this work consists of extending one-site mean-field
theory to larger blocks, following the idea of the Renormalization Group 
method. A fully variational method is obtained which incorporates 
correlations beyond mean-field at low computational cost and which is able,
unlike NRG, to accurately describe the different phases of the Mott-SF transition.
We judged the computational cost as
primordial, so that extensions to large lattices are in reach and so that a direct
simulation of the experimental parameters can be accomplished, while the
computational uncertainties remain well within the experimental uncertainty
range. A disadvantage of the method is the breaking of number
conservation during the intermediate steps of the renormalization
procedure. In the final step particle number should be restored in principle,
but this restoration is only partial when an insufficient amount of states are
kept during the model space truncation. The quantity that directly  relates to
experiment is the momentum distribution. It was argued in
Ref.~\cite{Kashurnikov02} that the appearance of satellite peaks in the
momentum distribution signals the appearance of a Mott region in the center of
the trap. We will especially focus on the momentum distribution and on the
issue of the satellite peaks.  

The organization of the paper is as follows: In Sec.~\ref{sec:Expla} we 
explain the method, in Sec.~\ref{sec:Comp} compare it to exact diagonalization
methods for small latices and in Sec.~\ref{sec:Resu} we present results in
1, 2 and 3 dimensions. We end with the conclusion and the acknowledgments.

\section{\label{sec:Expla}A Numerical Renormalization Group Method}

Bosons in an optical lattice realize a Bose-Hubbard
Hamiltonian~\cite{Jaksch98} and more specifically, we consider here the soft-core Bose-Hubbard
Hamiltonian in the grand-canonical ensemble in $d$ dimensions and subject to a
confining field,      
\begin{equation}
H = -t \sum_{\langle i,j \rangle} {b}^{\dagger}_i{b}_j +
 \frac{1}{2}U\sum_{i}{n}_i({n}_i-1)
 +\sum_{i} ( \epsilon_i - \mu) {n}_i.
\label{eq1}
\end{equation}
The sum $\langle i,j \rangle$ is over the nearest neighbors only. The
operator $b^{\dagger}_i$ creates a boson on lattice site $i$ while $b_i$
removes it. The operator $n_i$ counts the local density on site $i$. The
operator $N$ denotes the total number operator, $N = \sum_i n_i$. We take the
distance $a$ between adjacent sites equal to $a=1$. The 
confining field acts as a local site dependent chemical potential $\epsilon_i$
that can be added to the chemical potential $\mu$ to form $\mu_i = \epsilon_i -
\mu$. In case of a site dependent $\mu_i$ we speak of the inhomogeneous
or confined model, in case of a uniform $\mu$ we speak of the unconfined
or homogeneous model. We also define the coordination number $z = 2d$ as
the number of neighbors of each site and consider a linear, square or cubic
lattice of length $L$ along each axis. The energy scale is set by setting
$t=1$.   

This Hamiltonian is the easiest bosonic model in which two different effects
compete: the kinetic energy is diagonal in momentum space and tries to
delocalize the particles over the sites, while the potential energy is
diagonal in coordinate space and localizes the particles. 

\begin{figure}
\includegraphics[height=6cm]{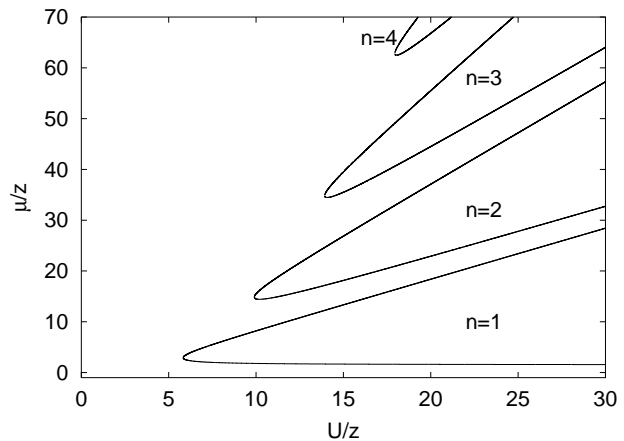}
\caption{\label{fig:phasemf}The mean-field phase diagram in the units we adopt
  throughout the paper. The Mott lobes are indicated and surrounded by a
  superfluid.}   
\end{figure}

We first discuss the physics of the model in absence of disorder in 
one dimension~\cite{Fisher89,Sachdev99}. When the potential energy
dominates, the system forms a Mott insulating phase at integer densities,
that remain pinned at these integer values and the phase is incompressible.
The Mott lobes are surrounded by compressible SF phases, where the densities
fluctuate. This is visualised in the mean-field phase diagram
Fig.~\ref{fig:phasemf}, which can easily be calculated~\cite{Fisher89,
Sachdev99}. Note that this phase diagram is approximate, in the true phase
diagram the $n=1$ lobe e.g. extends to smaller values of $U$ and the lobe
closes in a point-like fashion. There are two different phase transitions
possible. When keeping the density constant at an integer value, phase
fluctuations dominate and the transition  is of the
Berezinskii-Kosterlitz-Thouless (BKT) type. This transition can only occur at
the tip of the insulator lobe, that as a consequence closes in the point-like
fashion. The generic phase transition is driven by density fluctuations and
belongs to a different universality class. For a general $d$ dimension, the
BKT transition generalizes to the $(d+1)$ dimensional XY universality class.  

A first approximation for the Bose-Hubbard model is the Gutzwiller
variational ansatz~\cite{Sheshadri94,vanoosten01} leading to a
decoupling of the individual lattice sites and to a mean-field theory
(MF). This assumption can be described by the following substitution:  
\begin{equation}
 {b}^{\dagger}_i {b}_j \Rightarrow
     \psi_j {b}^{\dagger}_i + \psi^{*}_i {b}_j -  \psi_j\psi^{*}_i,
\label{eqMF}
\end{equation}
with $ \psi_i \equiv \langle {b}_i \rangle$. This leads to a model where particle
number symmetry can be broken, and that can exhibit a superfluid and a
Mott-insulator behavior.
 
In order to obtain a higher accuracy, we extend the MF approximation to a
a Renormalization Group procedure by taking more correlations into account. It
works as follows. Just as in MF, first break down the entire lattice to single
sites and solve the problem for each site separately. The Hilbert space is
truncated so that only a few basis states are kept on every site. In NRG this
state selection is based on energy solely, meaning that we keep the $N_s$
eigenstates corresponding to the $N_s$ lowest energy eigenvalues. The two
sites are combined now to form a small block. At this stage, the MF
approximation (~\ref{eqMF}) for the hopping term between the two sites can
be canceled by adding a term $(\psi_i^*-b_i^{\dagger})(\psi_j-b_j)$, after
which the Hamiltonian for the two sites becomes 
\begin{equation}
H_{12} = H_1 + H_2 + \sum_{i \in 1, j \in
 2}(\psi^*_i-b_i^{\dagger})(\psi_j-b_j) +  \textrm{h.c.}
\label{eqHnew}
\end{equation}
Here, $H_1$ and $H_2$ denote the Hamiltonians of the left and right site,
respectively, and the sum runs over adjacent sites that each belong to a
different block. In this first step, just the two sites $1$ and $2$ are
meant. The Hamiltonian $H_{12}$ is diagonalized in the space spanned by
the  product states, that are constructed from the individual basis states of
each site. After diagonalization, only a few states are kept again. Physical 
observables require now a rotation, since we have performed a basis
rotation. The procedure repeats itself: the small blocks can be joined to
form larger blocks which will themselves be the building blocks of still
larger blocks etc. This procedure is very similar to DMRG~\cite{White92}. 
The main differences are that in DMRG the
selection of the states is based on the eigenvalues of the density matrix
instead of on the lowest energy values, and secondly that in NRG one combines
blocks (exploiting symmetry), while in DMRG one extends the blocks site by
site. In NRG one performs one calculation till the lattice is entirely built
up, while in DMRG one sweeps again through the lattice till convergence is
obtained. DMRG yields results with a higher accuracy, but its computational
time and memory cost requirements are beyond current computer power for
dimensions higher than one.  

A new idea is that we improve on the standard NRG
procedure by adding source terms to the Hamiltonian on the edges of the
blocks. These terms compensate for the interaction with the other blocks in a
mean-field way. In this way, the Hamiltonian of the local block feels already
an average contribution of the blocks that have not yet accounted for, and
that have a non-local influence on the block under consideration.  After the two
joining blocks are taken together, these terms need to be extracted again. If
it were possible to work in an infinite Hilbert space the net effect of these
source terms would be zero, but in a truncated space the calculation will
depend on the values of these terms. E.g. suppose we are looking for a Mott
phase and these source terms are set to finite values, then we will not find
the Mott phase if the source term yields contributions to states higher than
the cut-off. This surely will be the case near the boundary of the Mott lobe
in a homogeneous system. We have also tried to apply improved periodic
boundary conditions by use of such source terms, contrary to DMRG where one
usually adopts open boundary conditions. (In general, periodic boundary
conditions are easier for finite-size scaling.)  We will come back to the
issue of source terms in the next section.    

Some operators, such as the total number operator squared $N^2$ also acquire a
contribution from the cross-terms between the two building blocks. So, more
than a simple rotation is needed in this case, and contributions from the
total number operator $N$ must be taken into account. Schematically, $\langle
N^2_{12} \rangle = \langle N_1^2 \rangle + \langle N_2^2 \rangle + 2 \langle
N_1 N_2 \rangle$, in which the indices $1,2$ indicate the two joining blocks. 

One can obtain the one-body density matrix in coordinate space with the MF+NRG
method, but at a low accuracy because correlations $b^{\dagger}_ib_j$ are 
inaccurate when $i$ is not the first site of the renormalization
procedure, and we need the entire matrix for a confined system. However, we
are able to directly calculate the diagonal of the momentum density operator
$\rho_{k,k} \equiv \rho_k$ (in one dimension), 
\begin{eqnarray}
\rho_{k} & =&  \sum_{i,j} e^{-ik(i-j)} b^{\dag}_i b_j \nonumber\\
{} & = & \sum_{i \in L, j \in R} \cos (k(i-j))b^{\dag}_i b_j.
\end{eqnarray}
Here we sum over all sites of the blocks, easing the problem encountered with
the one-body density matrix in coordinate space. This operator can be rewritten as
\begin{eqnarray}
{} & =&   \sum_{i \in L, j \in R}\cos (ki) \cos(kj) b^{\dag}_i b_j  + \sin(ki)
\sin(kj) b^{\dag}_i b_j \nonumber  \\
 {} & =&   \left( \sum_{i \in L}\cos (ki) b^{\dag}_i \right) \left( \sum_{j \in R}\cos
  (kj) b_j \right) \nonumber \\
{} & {} &  +  \left( \sum_{i \in L}\sin (ki) b^{\dag}_i \right) \left(
  \sum_{j \in R}\sin (kj) b_j \right) \nonumber \\
{} & = & C_{k_L}^{\dag} C_{k_R} + S_{k_L}^{\dag}S_{k_R}. 
\label{eq:momcossin}
\end{eqnarray}
Hence we have to keep track of linear combinations of the creation
$c^{\dagger}_i$ and annihilation $c_i$ operators. The parts in which both 
sites $i$ and $j$ belong to the same block (left(L) or
right(R)) have been omitted, since their updating consists only of a rotation
to the newly truncated basis. When the sites belong to different blocks, there
is also a contribution of the cross-term just as with the operator
$N^2$, but it still suffices to update the $C$ and $S$ operators. The
extension to higher dimensions of Eq.(\ref{eq:momcossin}) is
straightforward. We normalize the Fourier transform by adding prefactors
$\frac{1}{L}$ so that the trace of the density matrix in momentum space yields
the number of particles in the system. 

\section{\label{sec:Comp}Comparing the Method with Exact Results for Small
  Lattices} 
In this section we consider a small lattice in one dimension and in the
absence of any kind of disorder.

We have checked the code by comparing the resulting energies to direct Lanczos
diagonalization values for a lattice containing 8 sites in one dimension. The one
dimensional Bose-Hubbard model with periodic boundaries is a worst-case
scenario for our MF+NRG procedure. The
results are summarized in Table~\ref{table:lanc}. The parameters in the table vary
from a SF phase to a Mott phase. As expected, very deep in a Mott phase or in
a SF phase we obtain a very good accuracy. Note that the Lanczos
diagonalization was performed with a fixed boson number, while in the MF+NRG we
adjusted the chemical potential in order to fix the density. The deviations
should be interpreted accordingly. 

\begin{table}
\caption{\label{table:lanc}Comparison of the energies ($E_{16}$ and $E_{32}$)
  per site obtained by the MF+NRG method for a modest number of states (16 and
  32) kept after each diagonalization with the results $E_L$ of a Lanczos 
  diagonalization procedure for a small lattice of 8 sites in 1 dimension. The
  deviations ($D_{16}$ and $D_{32}$) are indicated and can be made smaller by
  keeping more states after each diagonalization. The mean-field values are in
  the last column.} 
\begin{ruledtabular}
\begin{tabular}{cccccccccc}
 $U$ &$\mu$ &$E_L$ &$E_{16}$ &$D_{16} (\%)$ & $E_{32}$ &$D_{32} (\%) $ & $E_{MF}$\\
\hline
2.0  & -0.5 & -1.359 & -1.337 & 1.57 & -1.347 & 0.91 & -1.24\\
4.0  &  0.7 & -0.932 & -0.901 & 3.38 & -0.914 & 1.99 & -0.78\\
6.0  &  1.8 & -0.656 & -0.612 & 6.65 & -0.635 & 3.15 & -0.44\\
8.0  &  2.5 & -0.494 & -0.467 & 5.60 & -0.483 & 2.40 & -0.27\\
10.0 &  3.5 & -0.397 & -0.382 & 3.53 & -0.392 & 1.09 & -0.10\\
12.0 &  4.0 & -0.331 & -0.324 & 2.35 & -0.330 & 0.31 & 0.00\\
14.0 &  5.0 & -0.284 & -0.283 & 0.50 & -0.284 & 0.23 & 0.00\\
16.0 &  6.0 & -0.249 & -0.247 & 0.38 & -0.248 & 0.07 & 0.00\\
18.0 &  7.0 & -0.221 & -0.221 & 0.30 & -0.222 & 0.04 & 0.00\\
20.0 & 15.2 & -0.199 & -0.181 & 8.70 & -0.199 & 0.17 & 0.00\\
\end{tabular}
\end{ruledtabular}
\end{table}

\begin{figure}
\includegraphics[height=6cm]{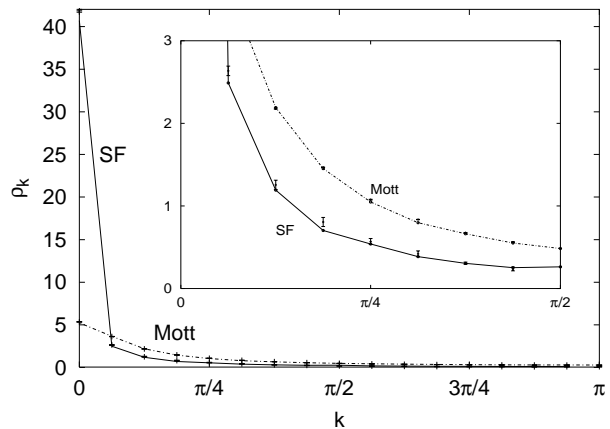}
\caption{\label{fig:momcomp}Checking the momentum distribution obtained with
  MF+NRG to a SSE calculation for an unconfined system in one
  dimension of 32 sites. Calculations have been done for
  a system in the Mott phase ($U = 6, \mu = 2$, dashed line) and for a system
  in the SF phase ($U = 2, \mu = 1$, full line). The errors on the SSE data
  points are shown but very small. In the inset, the MF+NRG data points are
  indicated explicitly by small circles, while "+" signs with error bars
  indicate the SSE data points.}   
\end{figure}

We have also checked observables like the local density $n_i$ and local
compressibility $\kappa_i$ against results obtained with the Stochastic
Series Quantum Monte-Carlo~\cite{Sandvik99} (SSE) method 
for larger lattices. Because the calculation of the momentum distribution
seems most critical, we have explicitly shown in Fig.~\ref{fig:momcomp} the
good agreement between the calculation of the momentum distribution with the
renormalization group and the SSE method for a SF and a Mott phase. The
one-body density matrix with the SSE method has been obtained by applying the
idea of Ref.~\cite{Dornreich01} to soft-core bosons. The SSE method served as
a testing ground for the renormalization method here. So we have shown that a
lot of physics might be examined with our new method. 

\begin{figure}
\includegraphics[height=6cm]{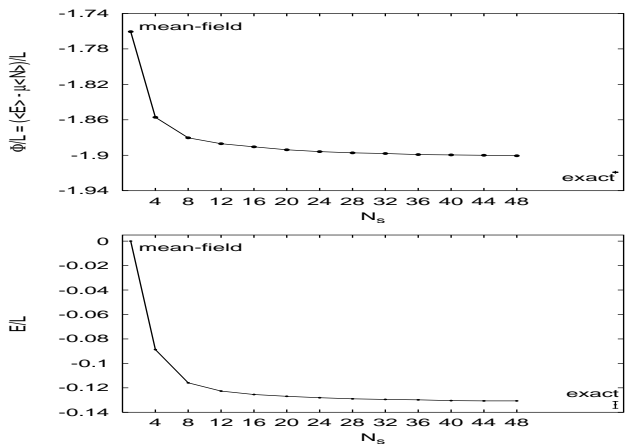}
\caption{\label{Fig:nsd} Upper: The system is variational in the average grand-canonical
  potential per site $\Phi/L = (\langle E \rangle - \mu \langle N
  \rangle)/L$. As the number  of kept states $N_s$ is increased, the
  grand-canonical potential decreases. The full line is a guide to the
  eye. The parameters are chosen such that the system is near a generic phase
  transition on the SF side ($U=4, \mu = 1, L=1024$). This corresponds to the
  worst case scenario for our method. The mean-field ($N_s = 1$) and exact
  (corresponding to $N_s = \infty$) results are indicated. Lower: The
  exponential convergence of the energy is shown. The system
  is deep in the Mott phase with parameters $U=30, \mu=16$ and $L=1024$.}   
\end{figure}

\begin{figure}
\includegraphics[height=6cm]{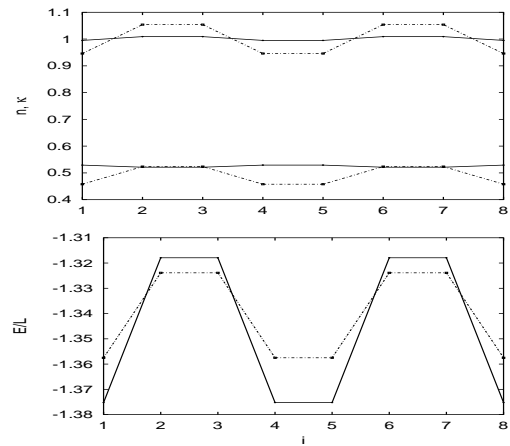}
\caption{\label{fig:brg} The figure shows how source terms can improve the
  calculations. Local energy per site $E_L$, local compressibility $\kappa_i$ and
  local density $n_i$ from bottom to top are plotted as a function of the
  lattice index $i$ for a homogeneous model of 8 sites in a SF phase ($U=2,
  \mu=-0.5$, the same values as the upper row in Table~\ref{table:lanc}). The
  dashed line has source terms set to zero, while for the full line they are
  set to their mean-field values.} 
\end{figure}

The parameter $N_s$ that fixes how many states are kept in the truncation
determines the accuracy of the results. As we have seen in
Table~\ref{table:lanc} energies can decrease by increasing $N_s$, while for
observables like the local density the fluctuations become smaller. We have
examined how the grand-canonical potential $\Phi$ decreases when $N_s$ is
increased in the upper part of Fig.~\ref{Fig:nsd} for a system of $L=1024$ sites (so that
finite-size effects can be filtered out of this discussion) in the SF phase
but very close to the generic Mott phase transition. This corresponds to the
worst case scenario for our method. Inclusion of just a few
states leads to a rapid decrease in the grand-canonical potential, but once
more than 20 states are kept, the potential decreases only very
gradually. The exact result $\Phi/L = -1.917(2)$ in Fig.~\ref{Fig:nsd} was
again obtained by the SSE-method, while with $N_s = 40$ we reached $\Phi/L =
-1.90$. Without source terms, we found that the calculated average
grand-canonical potential per site was $\Phi/L = -1.87$ with $N_s=40$, giving
further evidence of the usefulness of the source terms. It is the sweeping
property of the DMRG algorithm that could improve the results here
substantially, something we tried to avoid from the onset since this property
is computationally too costly in higher dimensions. The discrepancy with the
exact result in Fig.~\ref{Fig:nsd} reduces rather slowly at higher values of
$N_s$, primarily  because of the effects of block extension (reflected in the
curves of the local densities and local energies in  Fig.~\ref{fig:brg} for
the same effect) and complications due to the periodic boundary
conditions. The exact result should be recovered  in the limit of $N_s$ equal
to the dimension of the Fock space for each block. In addition, for a system
that is already deep in one of both phases, the MF+NRG method converges very
rapidly to the exact result, as the energy curve shows in the lower
part of Fig.~\ref{Fig:nsd} for a system in the Mott phase. Here, the energy
and grand potential differ only by a constant.  

On the other hand, the parameter $N_s$ also largely determines the required computer time: 
observables scale as $N_s^2$ per lattice site in memory cost, and the most
time consuming operation is the rotation of variables, which scales as
$N_s^6$ (multiplication of matrices of order $N_s^2$). All our calculations have been performed on a Pentium IV, 1.6GHz or a
Pentium III, each with 500MB RAM. Larger lattices and higher values of $N_s$
can straightforwardly be implemented on more performant hardware, but
requiring that all occupation numbers of the truncated states are arbitrarily
small on the one hand and on the other hand wishing to study large lattices in
high dimensions near a quantum phase transition is still not achievable.  

One of the crucial parameters of the method is the source term that is
inserted at the boundary of each block. If we set it to zero, our method 
reduces to the standard NRG method. It can be seen in Fig.~\ref{fig:brg}
that the fluctuations can be damped much better in a SF phase if we set the
source term equal to the MF expectation value of the operator $c_i$, while the
total energy deviates now $0.9 \%$ from to the exact result instead of $1.4\%$
without the source terms. For a homogeneous model in the thermodynamic limit,
the value of $\langle c_i \rangle$ should be site-independent and the source
terms should be chosen equal as well.  

As argued in the previous section, the source terms are optional and need to be chosen
carefully. It is well known~\cite{Sachdev99} that a Mott phase can only be
found if $[H, N] = 0$. Source terms might violate this condition near a
Mott-SF transition. The addition of source terms might lead to an incorrect
prediction of a SF phase when the compensation of the source terms in the
renormalization scheme is not complete. The source terms could yield contributions
to states that are thrown away after truncation of the Hilbert space and these
contributions can be quite large when the parameter $N_s$ is chosen too low. This 
might lead to an incorrect value of the transition point. In addition, even if
the transition point of a generic Mott-SF phase transition was known exactly,
and we would study the SF side of this transition, the source terms
should still be chosen carefully. This can be understood as follows: any net
contribution of a source term will deal with long-range correlations in the
same way as mean-field does, and we know that the
correlators predicted by mean-field are only valid in $d=3$ dimensions. E.g.
the density is a valid order parameter for the generic Mott-SF
transition~\cite{Sachdev99}, with  
\begin{eqnarray}
n \sim  & \mu^{1/2} & d=1 \nonumber\\
n \sim & \mu & d=3.
\label{eq:scaling}
\end{eqnarray}

\begin{figure}
\includegraphics[height=6cm]{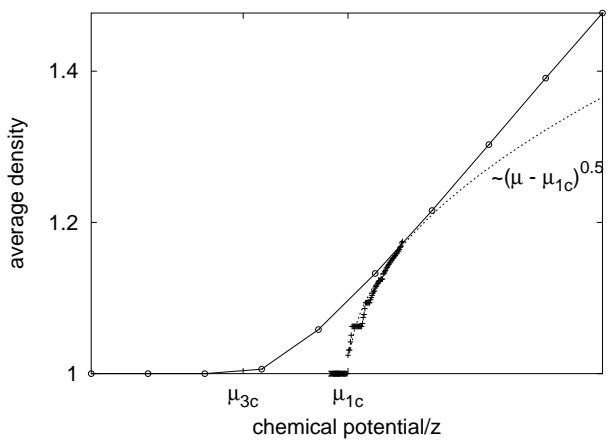}
\caption{\label{fig:nmu} Evolution of the density in the neighborhood of
  the generic phase transition between the Mott and SF phase for one ("+" marks) and
  three (empty circles) dimensions. The dashed line is a fit according to
  Eq.(~\ref{eq:scaling}). In these calculations the source terms are
  set to zero and lattices of size $L=1024$ sites were studied. Inclusion of
  the source terms in one dimension would lead to a similar plot as in the
  3D case.}  
\end{figure}

Briefly said, the source terms would in one and two dimensions lead to an
improved mean-field theory, in the sense that the correlators would
approximately have the same exponents as in mean-field theory and the Mott
lobe would extend a little bit farther into parameter space. This is also
explained in Fig.~\ref{fig:nmu} When $N_s$ is
large enough, these possible dangers become less severe. In the confined case,
any long-range correlations are effectively cut off and the addition of source
terms is always expected to improve the calculations, as has been verified. 

It was also tempting to study what happens if the blocks were extended
by a single site only. That leads to the same number of diagonalizations but
more rotations are needed. For the unconfined case, this yielded quite good
results, often smoother than in the Block Renormalization case. However, for
the confined case this procedure did not produce regions with integer density
and should hence only be used with great care.
 
\begin{figure}
\includegraphics[height=6cm]{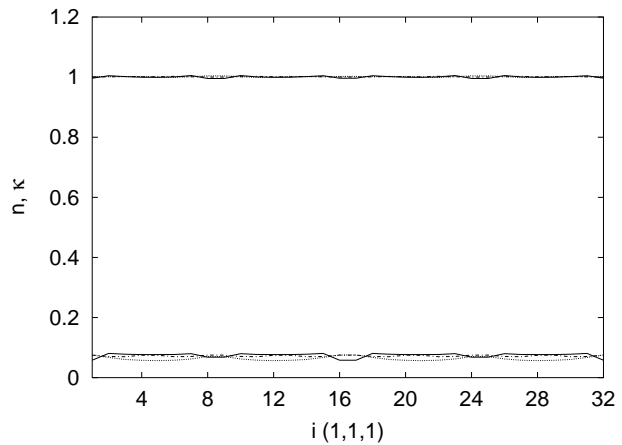}
\caption{\label{fig:etd_d} Local density (upper curves) and local
  compressibility (lower curves) for parameters $U/z=5.0$ and $\mu/z = 2.0$ in
  one (full line), two (dotted line) and three (dashed line) dimensions. MF
  theory predicts a SF phase, while the true phase in one, two and three
  dimensions is a Mott phase.} 
\end{figure}

The MF+NRG procedure also offers a substantial improvement over MF results. The MF
transition between the SF phase and the Mott phase is independent of the
dimension of the system and is located at $U_c \approx 5.83$, while a
DMRG~\cite{Kuhner99} study locates it at $U_c/z \approx 3.36$ in one 
dimension and a strong-coupling expansion~\cite{Elstner99} locates it at
$U_c/z \approx 4.18$ in two dimensions. We have performed a simulation at $U/z
= 5.0$ in one, two and three dimensions. While MF theory predicts a SF
phase (calculation yields $\langle c \rangle = 0.496$) the true phase should
be a Mott phase in one and two dimensions and we even found a Mott phase in
three dimensions. These results can be seen in Fig.~\ref{fig:etd_d} where the
local density and local compressibility are shown.

\section{\label{sec:Resu}Results}
Now that we have critically examined the approximations made in the MF+NRG, we
apply it to parameter regions where the results are unambiguous. In all
calculations the size of the lattices corresponds to the maximal achievable
size while a sufficient amount of states has been kept.

\subsection{\label{ssec:Res1}Results in one dimension}

We have, in complete analogy with Ref~\cite{Batrouni02},  confined a Bose gas
in a lattice of 128 sites in a trapping potential of the form
\begin{equation}
\epsilon_i = v_c (i - L/2)^2,
\end{equation}
with $v_c = 0.008$. Choosing the parameter this way allows for nice fillings
in the center and for densities going smoothly to zero near the edges of
the trap. 

\begin{figure}
\includegraphics[height=6cm]{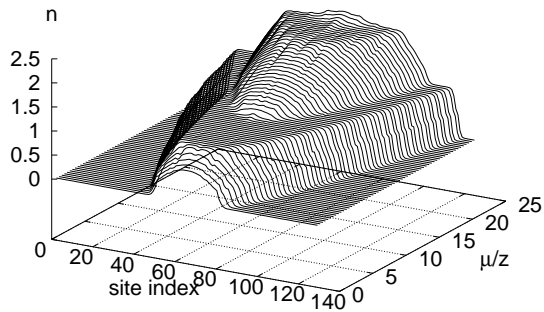}
\caption{\label{fig:d1_1n} Profile of the local density $n$ along the sites
  and as a function of the chemical potential $\mu$. The on-site repulsion is
  $U/2 = 7.1$ Above a certain value of $\mu$ we see the emergence of a plateau
  ($n = 1$), and when the total number of particles is even more increased, we
  see the re-emergence of a compressible region. This happens first around the
  center and continues to exist till a plateau with $n = 2$ is
  reached. These results confirm the result of Ref.~\cite{Batrouni02}}
\end{figure}

\begin{figure}
\includegraphics[height=6cm]{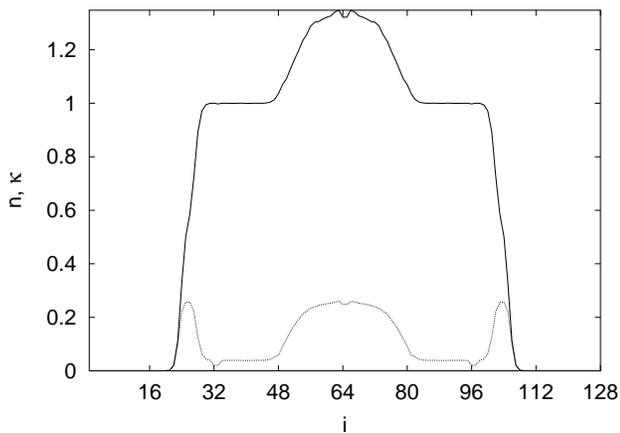}
\caption{\label{fig:d1_2nk} Profile of the local density $n$ (solid line) and local
  compressibility $\kappa$ (dashed line) along the sites for $U/2 = 7.1, \mu/2 = 6.1,
  L = 128, v_c = 0.008$, confirming again the result of Ref.~\cite{Batrouni02}.}
\end{figure}

In Fig.~\ref{fig:d1_1n} we see how plateaus with local fillings of an integer
number of bosons can arise as more and more particles enter the system. The
global compressibility is never zero, as it is the case in the unconfined
model in the thermodynamic limit. We cannot speak therefore of a true quantum
phase transition, the confining potential effectively cuts off all long range
correlations. However, in local regions the local compressibility can get very
low and the local  density can get stuck at integer values, reflecting a local
Mott region. This can be seen in Fig.~\ref{fig:d1_2nk}. All these results are
completely in line with those of Ref.~\cite{Batrouni02}. Also, for a canonical
calculation with an incommensurate filling a Mott phase with integer density
can still be found, because the confining potential changes the local
chemical potential. \\

Looking along the sites can be interpreted as different $\mu$-slices of the phase
diagram in the $(U, \mu)$ plane for the unconfined model~\cite{Fisher89}. This
allows to calculate the site at which a Mott domain is entered or left. It is
also clear that the BKT transition has no analog in the unconfined case. We
refer to Ref.~\cite{Batrouni02} for a state diagram. The authors of
Ref.~\cite{Batrouni02} also claim that $\kappa_i \sim (n_i-1)$ as the Mott
lobe is approached, independent of the on-site repulsion $U$ or the chemical
potential $\mu$. In our calculations the same behavior was seen for parameters
that are of the same order of magnitude, but for small and large values of $U$
the local compressibilities did not reach to the same values in the Mott region. 

\subsection{\label{ssec:Res2}Results in two dimensions}

\subsubsection{\label{sssec:Res2hom}Homogeneous case}

In principle it is possible to determine the phase diagram, but a complication
that makes a comparison more difficult is that in the
literature~\cite{Elstner99} calculations are usually based on a fixed density 
while we are working in the grand-canonical ensemble. The physically most
interesting case is with a disordered chemical potential~\cite{Prokofev03}. A
phase diagram requires a study of phase transitions and very close to a transition point it is
important to include more and more states into the truncated Hilbert space and
at the same time going to larger lattices. The method described here can only give a
qualitative answer and is not fit to quantitatively yield the exact location
of the transition point and does not allow to calculate the critical exponents
in an unambiguous way. \\

The problem encountered here is a 'memory effect' when $N_s$
is not high enough. When the source terms are set to zero and a large lattice
of $L = 256 \times 256$ is taken, a calculation with a too low $N_s$ predicts
a Mott phase while an increased $N_s$ leads to a SF phase. So, starting
from a Mott phase (zero source terms), results in a Mott phase and starting
form a SF phase (finite source terms) reveals a SF phase. The issue of the
phase diagram is very similar to the difficulties encountered with the strong-
coupling expansion by Freericks \textit{et   al.}~\cite{Freericks94}, although
their starting point is entirely different. As they point out, their method
cannot describe the physics close to the tricritical point, the density
fluctuations dominate even close to the tricritical point, and they can notice
that the shape of the Mott lobes has changed from one to higher
dimensions. Due to the limitations in our method we see the same qualitative
aspects, but we ran into the same quantitative difficulties, with the same
order of uncertainties. We will not report on calculations of the phase
diagram here.     

\subsubsection{\label{sssec:Re2con}Confined case}
\begin{figure}
\includegraphics[height=6cm]{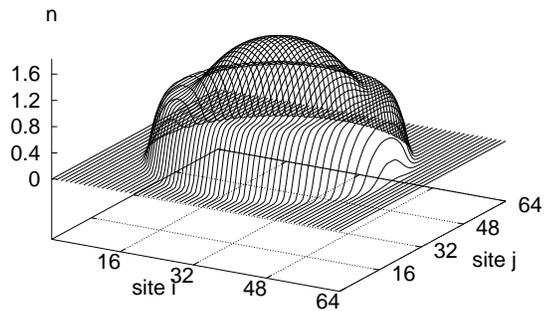}
\caption{\label{fig:d2_1n} Local density for a lattice consisting of $L = 64
\times 64$ sites and with parameters $U = 23.2, \mu = 28.0, v_c = 0.05$. Note
again the region with fixed integer density and the smooth transitions. }
\end{figure}

\begin{figure}
\includegraphics[height=6cm]{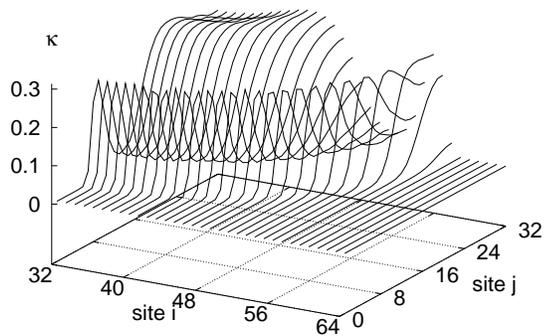}
\caption{\label{fig:d2_1k} Local compressibility as a function of the site
  indices and with the same parameters as in Fig.~\ref{fig:d2_1n}.}
\end{figure}

The trapping potential takes on each site $i$ the value 
\begin{equation}
\epsilon_i = v_cr_i^2,
\end{equation}
where $r_i$ measures the distance from the present site $i$ to the center of
the trap. The same holds in three dimensions.
In Fig.~\ref{fig:d2_1n} we plot the local density for a system of $L = 64
\times 64$ sites, with $U = 23.2, \mu = 28.0, v_c = 0.05$  and the space is constantly
truncated to 32 states, principally in line with our philosophy of a limited
but fast and reliable calculation. In Fig.~\ref{fig:d2_1k} we show the local 
compressibility for the same system, but only one quarter of the figure is
shown. The other parts are symmetric. Note again that there exists a Mott
insulating region with integer density. The transition from the SF region to
this Mott region is not sharp, and the local compressibility in the Mott
region is small but remains finite.

\begin{figure}
\includegraphics[height=6cm]{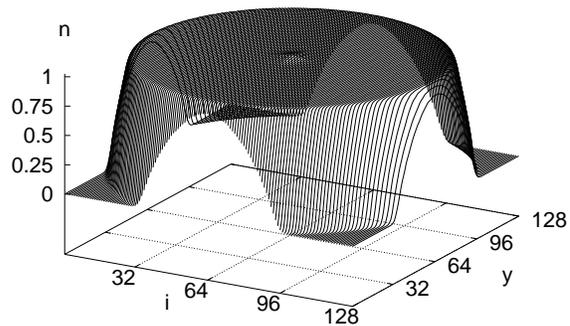}
\caption{\label{fig:d2_2n} Local density as a function of the site indices for
  a lattice of $L = 128 \times 128$ sites and with parameters $U = 22.0, \mu =
  35.6, v_c = 0.008$. The system is very close to developing a new SF peak in
  the center.}
\end{figure}

\begin{figure}
\includegraphics[height=6cm]{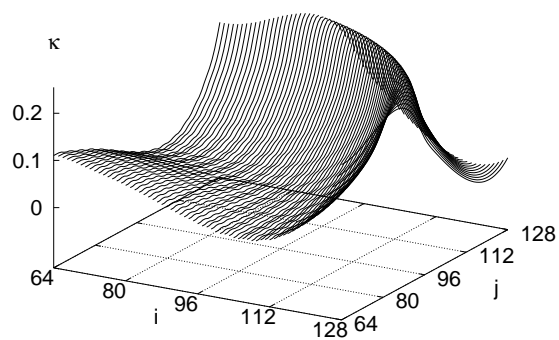}
\caption{\label{fig:d2_2k} Same as in Fig.~\ref{fig:d2_2n} but for the local
  compressibility now. The parts of the plot that are not shown are
  symmetric.}
\end{figure}

Another example can be found in Fig.~\ref{fig:d2_2n} and Fig.~\ref{fig:d2_2k}
for a lattice of $128 \times 128$ sites, showing Mott behavior and where for a
slightly weaker $U$ a new SF region would emerge in the center of the trap.

\subsection{\label{ssec:Res3}Results in three dimensions}
The original experiments by Greiner \textit{et al.}~\cite{Greiner02} were
performed in three dimensions, with laser beams cutting the atomic cloud in
about $L = 65 \times 65 \times 65$ sites, and a local density varying around
$n = 2.5$ atoms per 
site. Mott and SF behavior were demonstrated after examination of the
interference pattern of the laser images of the free expanding cloud. This
means that the
quantity of computational interest is the momentum distribution. In the
original  experimental setup, the absorption images of the three dimensional
distribution are taken along two orthogonal axes, revealing only the integral
over the third direction. The observed fading of the Bragg peaks had nothing
to do with the appearance of Mott behavior, and happened actually when the
system was already very deep in the Mott insulating phase. \\

\begin{figure}
\includegraphics[height=6cm]{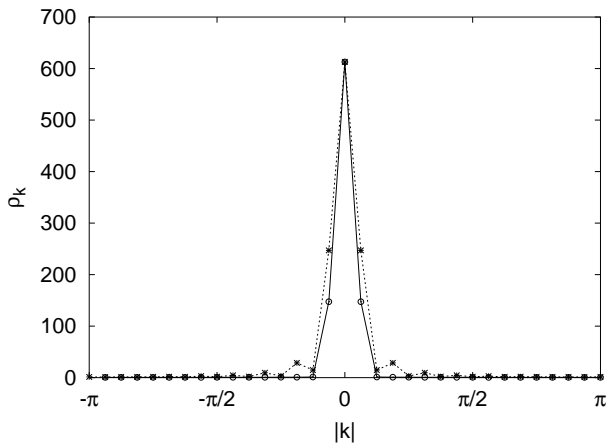}
\caption{\label{fig:d3_1}Momentum distribution for a system with a Mott plateau ($n=1$)
 in the center. The parameters are $U/z = 6.5, \mu/z=2.6, v_c = 0.04$ and $L = 32
  \times 32 \times 32$. The dashed line represents the distribution along the
  (1,0,0) direction, while the full line is taken along the (1,1,1)
  direction. According to Ref.~\cite{Kashurnikov02} the satellite peaks in the
  dashed curve point at an emerging Mott region in the center of the trap. } 
\end{figure}

So what could be a clear signal of the transition?  It was argued in 
Ref.~\cite{Kashurnikov02} that satellite peaks in the momentum distribution
are related to the appearance of a Mott region in the center of the
trap. Once the Mott region spanned almost the entire lattice, the peaks
disappeared into the typical broad, low-peaked Mott distribution. However, their
Worm Monte-Carlo calculation was only on a lattice of $L=16 \times 16 \times 16$ and it can be
expected that for a larger lattice the central SF peak might be so dominant
that the satellite peaks can hardly be resolved. We present a calculation
on a lattice of $L=32 \times 32 \times 32$. We show the momentum distribution
in Fig.~\ref{fig:d3_1} along the $(1,0,0)$ axis, for a system with an emerging
Mott region of $n=1$ in the center of the trap. As in
Ref.~\cite{Kashurnikov02} we see satellite peaks along the $(1,0,0)$ 
direction, but the central peak dominates. The satellite peaks are only 
$4.5\% $ in magnitude of the central peak and will be difficult to resolve in
practice. For the experimental setup with its lattice of about $L = 65 \times
65 \times 65$ sites, the situation will even be worse. We also note that the
satellite peaks depend on the direction of investigation, no satellite peaks
were seen e.g. along the $(1,1,1)$ direction in Fig.~\ref{fig:d3_1}. This
direction dependency is a consequence of the breaking of rotational symmetry
in a finite lattice, and its effects should diminish when larger lattices are
taken. 

\begin{figure}
\includegraphics[height=6cm]{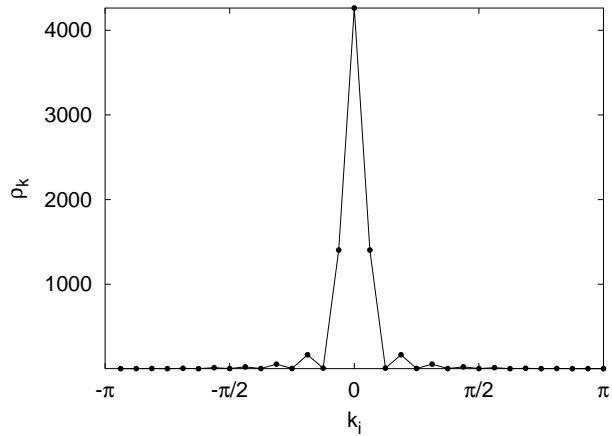}
\caption{\label{fig:mom23}Momentum distribution along the $(0,0,1)$ axis for a
  system with particle densities varying between $n=2$ and $n=3.2$ ($U/z = 11,
  \mu /z = 30, v_c = 0.1, L = 32 \times 32 \times 32$). The appearance of
  satellite peaks cannot be related to the \textit{emergence} of a Mott
  region in the center of the trap.} 
\end{figure}

Furthermore, the average density in the experiments was about $n \approx 2.5$
in the center of the trap~\cite{Greiner02}. There are no 
satellite peaks when the central density is non-integer, despite a broad Mott
$n=1$ region for an on-site repulsion $U$ that is strong enough. This Mott
region is reflected in the tail of the momentum
distribution~\cite{Kashurnikov02}. For a system with local densities varying
between $n=2$ and $n=3.2$ (all non-integer densities), we nevertheless found
satellite peaks in Fig.~\ref{fig:mom23}, and calculations showed the
same behavior for densities ranging between $n=3$ and $n=4$. These peaks
cannot possibly be related to the emergence of a Mott region in the center of
the trap. Local densities of $n=2$ at the border of the trap cannot occur
experimentally, but this situation can be thought of as the central region of
a larger lattice, from which the outer regions are not trapped any more.   

\begin{figure}
\includegraphics[height=6cm]{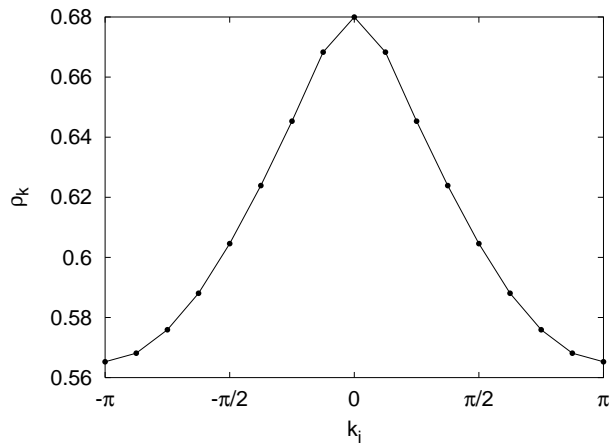}
\caption{\label{fig:mom03} Momentum distribution along the $(0,0,1)$ axis for
  a system with particle densities varying between $n=0$ at the edge of the
  trap to $n=3$ at the center ($U/z = 30, \mu/z = 75, v_c = 9.1,L = 16
  \times 16 \times 16$), leading to 2280 particles in the system. The
  distribution is broad and not peaked, signaling virtually overall Mott
  behavior. }   
\end{figure}

When going to higher values of
$U$ and $\mu$, it is in principle possible to have Mott phases at $n=2$
and $n=3$. In the mean-field phase diagram of the homogeneous
model~\cite{vanoosten01}, the different Mott lobes corresponding to densities
$n=1, n=2,$ etc. get closer to each other along the direction of the chemical
potential $\mu$ (see Fig.~\ref{fig:phasemf}). With the
confining potential $\epsilon_i$ present, the local densities along the
different sites can be interpreted as a scan of the homogeneous
model~\cite{Fisher89}. Hence in a small finite lattice it is not a priori
clear if there are non-integer densities between the different broad Mott
regions.
For a system with parameters $U/z = 30, \mu/z = 75, v_c = 9.1$ and $ L
= 16 \times 16 \times 16$, we found very few non-integer densities. The
density profile consisted of four plateaus with $n = 0,1,2,3$ respectively.
We have almost a superposition of four Mott phases leading to the
momentum distribution in Fig.~\ref{fig:mom03}, which is very low peaked and
broad. When the local particle density in the center of the trap is gradually
increased from $n < 3$ till the Mott region with $n=3$, and while there
already exist broad Mott regions with $n=1$ and $n=2$, we did not witness any
satellite peaks, because the Mott behavior of the $n=1$ and $n=2$ plateaus
already dominated the momentum distribution. \\

We are led to the observation that it will be difficult to indicate the
transition experimentally by satellite peaks, and that only
examination of the intensity and the width of the central peak along one
direction might be at hand to directly reveal Mott behavior.\\

\section{Conclusion}
In summary, we studied the Bose-Hubbard model subject to a confining
field in the grand-canonical ensemble. We combined the Gutzwiller mean-field (MF)
ansatz with a numerical (block) renormalization group method (NRG)
and we could calculate observables like local densities, energies,
the momentum distribution etc. The goal was to achieve variational results with energies much lower
than in mean-field theory and at a low computational cost in order to make
studies of large lattices in higher dimensions feasible. We have extensively
discussed the advantages and limitations of this new method.  The inclusion of 
source terms on the edges of the blocks improved results in the SF phase.\\

We have examined the smooth transition between SF and Mott regions in the
presence of a confining field. Although there is no real 'order parameter' to
be found in the momentum distribution, the momentum distribution can
nevertheless reveal important qualitative differences between pure SF systems
and systems with dominant Mott behavior. These differences can experimentally
best be seen in the central peak. Possible satellite peaks might be difficult to
resolve when the total number of confined particles is large and when the
filling factors are not of order unity. 

\section{Acknowledgments}
The authors wish to thank I. Bloch, M.Greiner, T.Papenbrock, J. Ryckebush and
D. Van Neck for valuable discussions. This work was supported by the Research
Board of the University of Ghent and the Fund for Scientific Research, Flanders.

\end{document}